%% file: sigproc-sp.tex
\documentclass{sig-alternate}

\usepackage{url}
\usepackage{multirow}

\usepackage{array}
\usepackage{rotating}
\usepackage{algorithm}
 \usepackage[noend]{algpseudocode}
\usepackage{color}
\usepackage{psfrag}
\usepackage{makecell}
\usepackage{wrapfig}
\usepackage[framemethod=tikz]{mdframed}
\usepackage{lipsum}
\usepackage{varwidth}

\newenvironment{WrapText}[1][r]
  {\wrapfigure{#1}{0.28\textwidth}\mdframed[backgroundcolor=green!20,skipabove=0pt,skipbelow=0pt]}
  {\endmdframed\endwrapfigure}

\begin{document}
\makeatletter
\let\@copyrightspace\relax
\makeatother

%\title{Stay where you belong: on the permanence of vertices in network communities}
\title{On the categorization of scientific citation profiles in computer sciences}

\numberofauthors{1} 
\author{\alignauthor *Tanmoy Chakraborty$^{1}$, Suhansanu Kumar$^{2}$, Pawan Goyal$^{3}$, \\ Niloy Ganguly$^{4}$, Animesh Mukherjee$^{5}$ \\
\affaddr{Department of Computer Science \& Engineering, \\Indian Institute of Technology, Kharagpur, India -- 721302} \\
\email{\{$^1$its\_tanmoy, $^2$suhansanu.kumar, $^3$pawang, $^4$niloy, $^5$animeshm\}@cse.iitkgp.ernet.in }\\
({\bf {\color{blue} Accepted in Communications of the ACM (CACM)}})
}

\maketitle
\begin{abstract}
A common consensus in the literature is that the citation profile of published articles in general follows a universal pattern -- an initial
growth in the number of citations within the first two to three years after publication followed by a steady peak of one to two years and
then a final decline over the rest of the lifetime of the article. This observation has long been the underlying heuristic in determining
major bibliometric factors such as the quality of a publication, the growth of scientific communities, impact factor of publication venues
etc. In this paper, we gather and analyze a massive dataset of 1.5 million scientific papers from the computer science domain and notice that {\em the citation count of the articles over the years follows a remarkably diverse set of patterns} -- a profile with an initial peak (PeakInit), with distinct multiple peaks (PeakMul), with a peak late in time (PeakLate), that is monotonically decreasing (MonDec), that is monotonically increasing (MonIncr) and that can not be categorized into any of the above (Oth). 
\begin{WrapText}
{\bf Key insights:}\\
 $\bullet$ {\color{blue}Analyzing a massive dataset of computer science domain reveals six distinctive citation trajectories of scientific articles.}\\
 $\bullet$ {\color{blue}After suitable characterizations of these profiles, major modifications of the existing bibliographic indices seem to be a compelling task.}\\
 $\bullet$ {\color{blue}Unlike existing network-growth models, these trajectories can only be reproduced once both ``preferential attachment'' and ``aging'' are taken into account together.}  
\end{WrapText}
We conduct a thorough experiment to investigate several important characteristics of these categories such as how individual categories attract citations, how the categorization is influenced by the year and the venue of publication of papers, how each category is affected by self-citations, the stability of the categories over time, and how much each of these categories contribute to the core of the network. Further, we show that the traditional preferential attachment models fail to explain these citation profiles. 
Therefore, we propose {\em a novel dynamic growth model} that takes both the {\em preferential attachment} and the {\em aging factor} into account in order to replicate the real-world behavior of various citation profiles. We believe that this paper opens the scope for a serious re-investigation of the existing bibliometric indices for scientific research.

\end{abstract}

\input{introduction.tex}

\input{result.tex}

\input{discussion.tex}

\newpage

\section{Supporting Information}

\subsection{Dynamic Growth Model}\label{model}
Unlike the standard growth models proposed previously for citation networks, our model takes into account
both ``preferential attachment''
\cite{Albert2002,  Price} and ``aging'' \cite{RePEc:eee} of each paper in order to
synthesize the network. To begin with, we include the first six years (1970 - 1975 (inclusive)) network information from the real
dataset to bootstrap the model. This information includes -- induced
subgraph of the papers published within these years, their category
information and the year of publication. Apart from this, for fair comparison with the real-world dataset, we use two more distributions
generated from the real-world dataset -- year-wise distribution of the number of publications (denoted as {\em Pub-dist}) and the reference
distribution (i.e., number of references vs. fraction of papers having those many references, denoted as {\em Ref-dist}). These form the base
statistics of our model.

Since we know the category information of the papers which are considered in the base statistics, we form six buckets corresponding to the six categories. Each bucket
constitutes papers of the corresponding category. Now for inserting new papers in the network at each time step $t$ (corresponds to a
particular
year), we execute the following steps in order:\\
1. Select the actual number of papers (say, $N$) published at time $t$ from {\em Pub-dist}, and create a set of $N$ number of nodes (say,
$P_N$)
    to be inserted in the network.\\
2. For each such $p$ present in $P_N$:

\hspace{2 mm}  2.1 Select one of the buckets (categories) for $p$ preferen-

\hspace{8 mm}tially based on the size of the buckets.

\hspace{2 mm}  2.2 Select a value $R$ from {\em Ref-dist} to determine the num-

\hspace{8 mm}ber 
of outward edges (references) emanating from $p$.

\hspace{2 mm}  2.3 For each outward edge, select the other end-point 

\hspace{8 mm}using the following steps:

\hspace{7 mm}  2.3.1  Select a bucket $B$ preferentially based on the 

\hspace{16 mm}number of incoming citations obtained by the 

\hspace{16  mm}papers in different
buckets till time $t-1$.

\hspace{7 mm}  2.3.2  Select one of the papers from $B$ based on its 

\hspace{16 mm} {\em attractiveness} at time $t-1$.

\hspace{2 mm}  2.4 Finally, assign $p$ in the selected category.

\noindent The {\em attractiveness} ($\pi_i$) of a paper $p_i$ is determined by the category where it falls using the two factors -- the time
after
its publication (aging) and the total citation count it has accumulated so far (preferential attachment) as follows. The notations used to
describe the formulation of $\pi_i$ for the different categories are -- $k_i$: the in-degree of paper $p_i$, $\mu$: average citation count of papers present in the
category where $p_i$ belongs to,
$\rho$: a parameter of the 
model used to dampen certain factors associated with it, $\tau$: a
%PG: \rho definition not clear
parameter of the model used to scale the time of occurrences of the peaks.

\begin{itemize}
 \item \textbf{MonDec:} Each paper declines in its popularity exponentially with time but remains proportional to its overall
significance, i.e., 
\vspace{-2mm}
\begin{center}
 $ \pi_{i} = \frac{k_{i} + \mu} { \exp ({\rho t}) }; \ \forall t $
\end{center}

\begin{figure*}[ht!]
 \begin{center}
\scalebox{0.35}{
\includegraphics{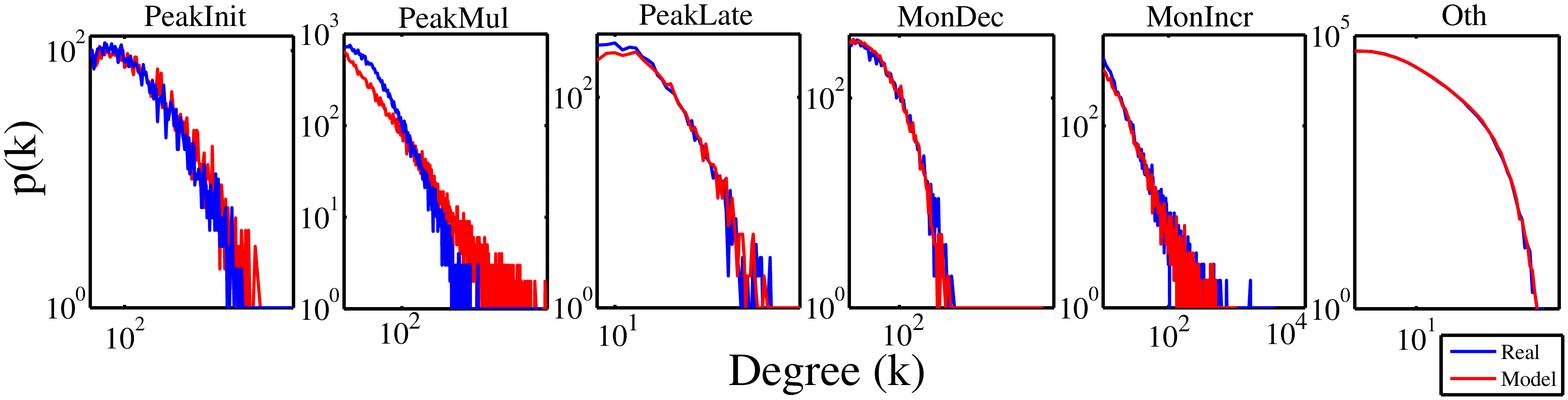}}
\end{center}
\caption{{\bf Figure S1.} Comparison of the in-degree distributions obtained from real-world dataset and the model  separately for
 different categories (both the axes are in log-scale).}\label{deg_eval}
\end{figure*}

\item \textbf{PeakInit:} We assume $T$ as the occurrence of the early peak in such distribution. PeakInit pattern comprises two phases as
commonly hypothesized for most of the earlier papers -- initial increase through preferential attachment and then a continuous decline.
Therefore,
\vspace{-2mm}
\begin{center}
 $\pi_{i} = k_{i} + \mu;\ if\ t \leq T + \tau$\\
$\pi_{i} = \frac{k_{i} + \mu} { \rho t } ; \ otherwise $
\end{center}

\item {\bf PeakLate:}  We assume $T$ as the occurrence of the late peak in such distribution. PeakLate behaves similarly as PeakInit with
the only difference that it experiences its peak at a much later period after the time of publication. Therefore,
\vspace{-2mm}
\begin{center}
$\pi_{i} = k_{i} + \mu;\  if\ t \leq T+\tau$\\
$\pi_{i} = \frac{k_{i} + \mu} { \rho t } ;\ otherwise $
\end{center}

\item \textbf{PeakMul:} We empirically observe that about 65.04\% papers in PeakMul category have two peaks on an average in the timeline of the
citation profile. Therefore, in this model we assume that $T_1$ and $T_2$ are
the times of two peaks respectively. The pattern of this profile can be
sub-divided into four phases -- initial rise through preferential attachment followed by a
monotonic decline, a second rise through preferential attachment followed by a monotonic decline. Therefore, 
\vspace{-2mm}
\begin{center}
$\pi_{i} = k_{i} + \mu ; \  if\ 0 \leq t \leq T_1+\tau$\\
$\pi_{i} = \frac{k_{i} + \mu} { t } ;\  if \ T_1+\tau < t \leq \frac{T_1+T_2}{2}+\tau$\\
$\pi_{i} = k_{i} + \mu ;\ if \ \frac{T_1+T_2}{2}+\tau < t \leq T_2+\tau$\\
$\pi_{i} = \frac{k_{i} + \mu} { t };\ otherwise $
\end{center}

\item \textbf{MonIncr:} In this category, each paper acquires popularity directly proportional to its earlier citation count as well as
time. Thus,
\vspace{-2mm}
\begin{center} 
$\pi_{i} = k_{i} + \rho \mu + t   ;\  \forall t $\\
\end{center}

\item \textbf{Oth:} Since the papers in this category do not follow any particular pattern of citation profile, we do not add the aging
factor here. When a new added paper tries to connect the references to the papers in this category, the papers are selected uniformly at
random, i.e., the $\pi$ values for all the papers in this category are same and remain constant over time.

\end{itemize}
The values of $\pi$ per paper and the value of $\mu$ for each category are calculated on the fly at each time $t$ based on the information at time
$t-1$. The time of occurrences of peaks ($T$, $T_1$ and $T_2$) are selected from the actual distribution of peak occurrence
time for each category individually. The results in Figure 3 (bottom panel) of the main text are an outcome of 100 simulation averages. The final values of the
model parameters determined in order to closely match the simulation results with the real-world patterns are as follows: $\tau$ = 1
(PeakInit) and =3 (rest); $\rho$ = 0.25 (MonDec), =0.7 (PeakInit), =0.5 (PeakLate) and =0.3 (MonIncr).

\begin{figure}[!h]
\centering
\includegraphics[width=0.7\columnwidth]{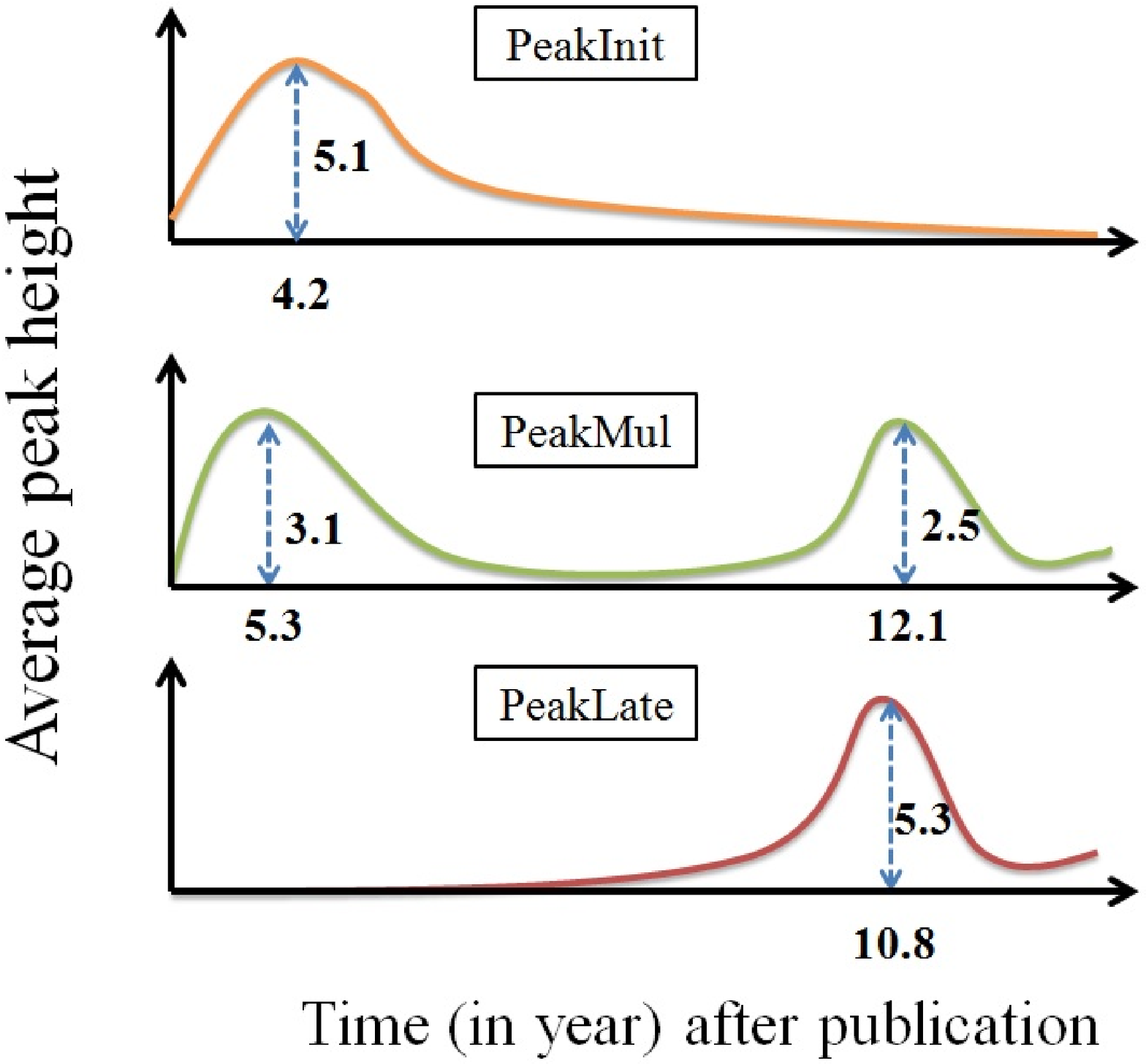}
\caption{{\bf Figure S2.} Hypothetical lines showing number of peaks, average height and average time of occurrences of the peaks for
PeakInit,
PeakMul and PeakLate categories.}\label{intermediate}
\end{figure}

\subsection{Comparing the Emergent Degree Distributions for Different Categories}

In the earlier section, we have proposed a model that replicates different citation patterns observed in real-world dataset.
In Figure 3 (bottom panel) of the main text, we have observed that the results obtained from our dynamic model have a significant resemblance with the
citation profiles observed in the real data. We further evaluate our model in terms of degree distribution. Since we use
out-degree distribution (reference distribution) in our model as an input, we evaluate the model with respect to the in-degree
distribution. Figure S1 shows the in-degree distributions obtained from the model and the real-world dataset separately for
different categories. We observe that the outputs of the model have a significant resemblance with the real-world results. This in turn further strengthens the applicability of our model to reproduce the real-world phenomenon.

\subsection{Intermediary Behavior of PeakMul Category} 
PeakMul seems to have a significantly unique citation profile in comparison to all the other categories. For instance, contrary to the
other categories where the profile is mostly determined by the height and the
time
of peak occurrence, the characteristic of this category is controlled
by one more parameter -- the number of peaks. Therefore for the PeakMul category shown in Figure 3 of the main text, the line
depicting the average behavior (red line) does not have much resemblance with the corresponding representative instance (broken black
line). In
Figure S2, we show the average height and the average time of occurrences of the peaks for PeakInit,
PeakMul and PeakLate categories. A deeper analysis unfolds three interesting observations -- (i) most of the papers (65.04\%) in
PeakMul
category have two peaks on an average in the timeline of the citation profile, (ii) the sum of the average heights of first two
peaks in PeakMul
category (i.e, $3.1+2.5=5.6$) is (nearly-)similar to the height of the peak for PeakInit and PeakLate categories ($5.1$ and $5.3$
respectively), (iii) the average difference between the time of occurrences of the first two peaks in PeakMul category (i.e.,
$12.1-5.3=6.8$) is
(nearly-)similar to the difference of the occurrence of the peak in PeakLate and PeakInit categories (i.e., $10.8-4.2=6.6$). From these observations,
one could argue that PeakMul behaves like the intermediary between PeakInit and PeakLate categories. We have used these observations in
order to configure the dynamic growth model described in Section \ref{model}. The detailed analysis of this category remains as one of the potential
areas of future research.

\bibliographystyle{abbrv}
\bibliography{ref}  % sigproc.bib is the name of the Bibliography in this case

\end{document}

%% file: introduction.tex
\section{Introduction}
Quantitative analysis in terms of counting, measuring, comparing quantities and analyzing measurements is perhaps the main tool to
understand
the impact of science. With the progress of time, scientific research itself by recording and communicating research results through scientific
publications, has become enormous and complex. The complexity has become so specialized that individual understanding and experience are no
longer sufficient to unfold trends or for making crucial decisions. 
Therefore, an exhaustive
analysis of research outputs in terms of scientific publications is of great interest among scientific communities to be selective, 
\begin{WrapText}
{\color{blue} According to Eugene Garfield, a {\bf citation} is nothing but a means
to
(i) pay homage to pioneers, (ii) give credit for related work (homage to peers), (iii) identify methodology, equipment etc., (iv)
provide background reading, (v) correct one's own work or the work of others and so on. }
\end{WrapText}
to highlight significant or promising areas of research, and to manage better investigation in science \cite{meho, Peritz, Boonchom2012220, Radicchi:2008p6887}.
Bibliometrics (aka Scientometrics) \cite{Bensman, Vincent}, the application of
quantitative analysis
and statistics to publications such as
research
articles and their accompanying citation counts, turns out to be the main tool for such an investigation.
From the pioneering
research of  Garfield \cite{garfield}, the use of citation analysis in bibliographic research serves as the fundamental quantifier for evaluating the contribution of researchers and research outcomes.

\begin{figure}[!h]
\centering
\includegraphics[width=\columnwidth]{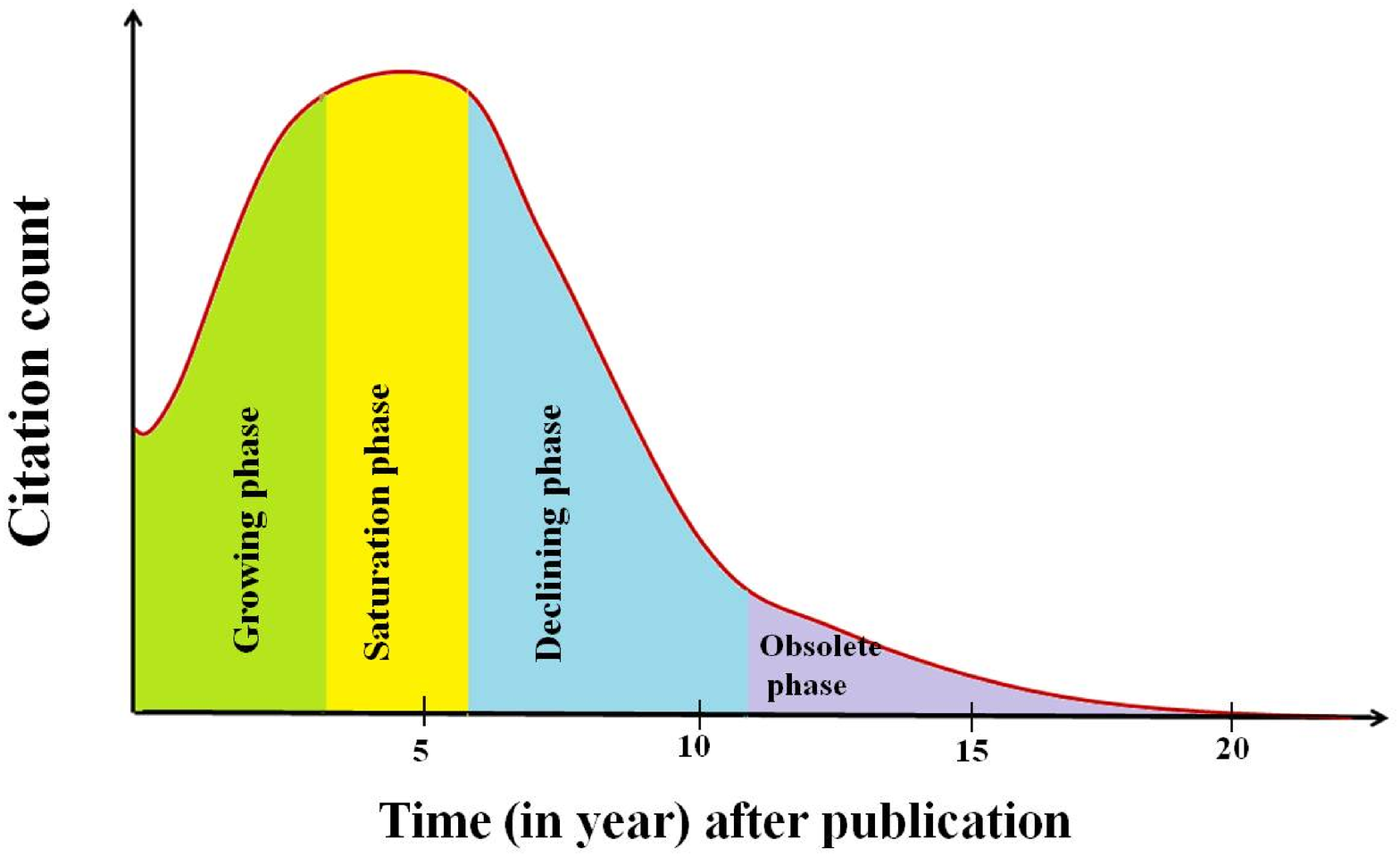}
\caption{(Color online) A hypothetical example showing the traditional belief in the pattern of citation profile of a scientific paper after publication.}\label{belief}
\end{figure}

Citation network represents the knowledge graph of science where individual papers are knowledge sources and their interconnectedness in terms of citation 
represents the relatedness among various kinds of knowledge. For instance,  a citation network is considered to be an effective proxy for studying disciplinary knowledge flow, is used to discover knowledge backbone of a particular research area, and helps in grouping similar kinds of knowledges and ideas.
Numerous research have been conducted on citation networks and their evolution over time. There is already a well-accepted belief about the dynamics of citations that a scientific article receives
after publication -- an
initial growth (growing phase) in the number of citations within the first two/three years after publication followed by a steady peak of
one to two years (saturation phase)
and then a final decline over the rest of the lifetime of the article (decline and obsolete phases) as shown in Figure \ref{belief}
\cite{garfield2, garfield1, garfield_06}. In most cases,
the above observation has been drawn from the analysis of a very limited set of publication data \cite{Callaham, garfield_86}, thus,
obfuscating the true characteristics. Here, we conduct our experiment on a massive bibliographic dataset of the computer science domain
comprising more than 1.5 million papers published between 1970 and 2010.  {\em Strikingly, unlike earlier observations
about citation profile of a paper, we notice six different types of citation profiles} prevalent in the dataset (namely, PeakInit,
PeakMul, PeakLate, MonDec, MonIncr and Oth). 
\begin{WrapText}
 $\bullet$ {\color{blue}Earlier, citation trajectory of an article was assumed to be increasing initially and then following a downward growth.}\\
 $\bullet$ {\color{blue}We observe six distinct citation trajectories after analyzing a massive dataset of computer science domain.} \\
 $\bullet$ {\color{blue} Since the citation profile can be categorized into at least six different types, all measures of scientific impact (e.g., impact factor) require a serious revisit.}
\end{WrapText}
We exhaustively analyze these
profiles to exploit the micro-dynamics controlling the actual growth of the underlying citation network that has remained unexplored in the
existing
literature. This categorization allows us to {\em propose a holistic view of the growth of citation network through a dynamic model} that takes into account the well-accepted concept of {\em preferential attachment} \cite{Albert2002, Barabasi99, Price} along with the {\em aging factor} \cite{RePEc:eee} of scientific articles in order to reproduce different citation profiles observed in the real-world dataset.
To
the best of our knowledge, this is the first attempt to consider these two factors together in synthesizing the dynamic growth process of
citation profiles. 
We believe that the key observations made in this paper will not only help in reformulating the existing bibliographic indices such as Journal Impact Factor (JIF), but
will also enhance the general bibliometric research such as citation link prediction, information retrieval and
self-citation characterization.

%% file: result.tex
\section{A Massive Publication Dataset}
Most experiments in the literature on analyzing citation profiles have worked with small datasets. However in this experiment, we gather and
analyze a massive dataset to validate our hypothesis. We have crawled one of the largest publicly available datasets from Microsoft Academic
Search (MAS)\footnote{\url{academic.research.microsoft.com}} which houses over 4.1 million publications and 2.7 million authors with updates
added every week~\cite{asonam}. We collected all the papers published in the computer science domain and indexed by
MAS\footnote{The crawling process took around six weeks and completed in August, 2013.}.  The crawled dataset contains more than 2 million
distinct papers altogether which are further distributed over 24 fields of computer science domain (as categorized by MAS).
Moreover, each paper comes along with various bibliographic information -- the title of the paper, a unique index for the paper, its
author(s), the affiliation of the author(s), the year of publication, the publication venue, the related field(s) of the paper, the
abstract and the keyword(s). 

In order to remove the anomalies that crept in due to crawling, the dataset was passed through a series of initial preprocessing stages. The filtered dataset contains more than 1.5 million papers with 8.68\% papers belonging to multiple fields (act as interdisciplinary
papers). We have made the dataset publicly available at \url{http://cnerg.org} (see ``Resources'' tab).

\begin{figure}
 \begin{center}
\includegraphics[width=0.35\textwidth]{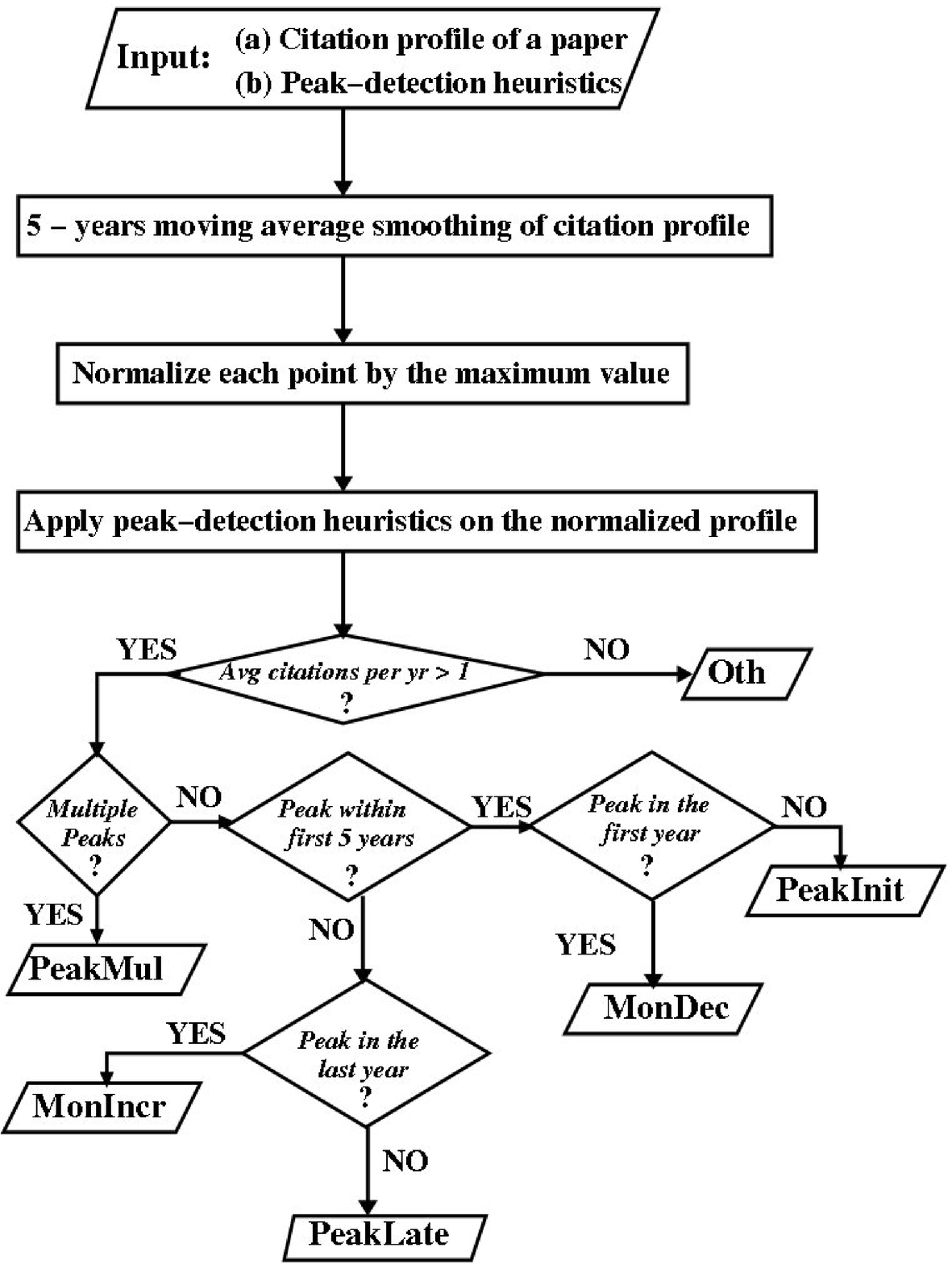}
\end{center}
\vspace{-5mm}
\caption{A systematic flowchart demonstrating the rules for classifying the training samples.}\label{algo}
\end{figure}

\begin{figure*}[t!]
\centering
\scalebox{0.35}{
\includegraphics{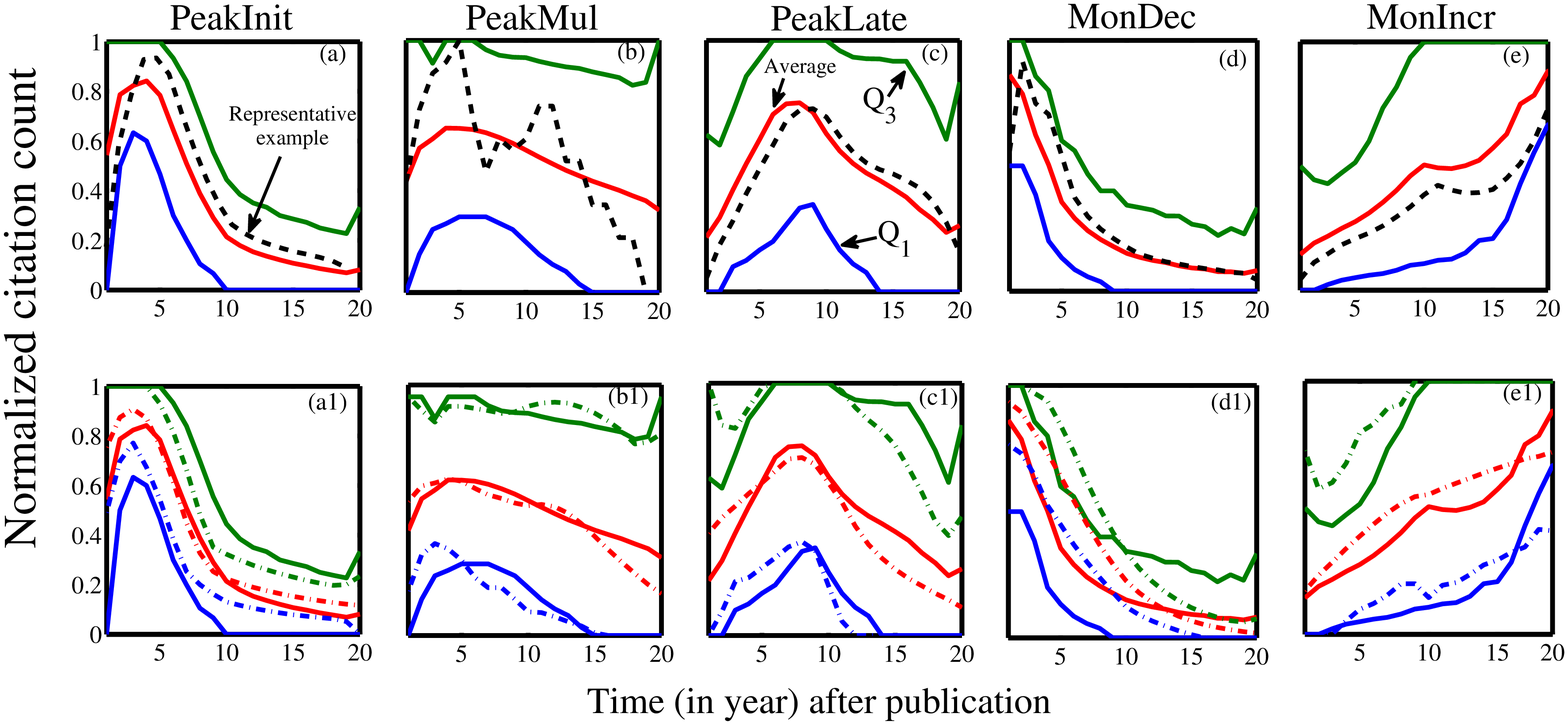}}
\caption{(Color online) Citation profiles for the first five categories obtained from analyzing the real-world citation dataset (top panel)
and a comparison of that with the results obtained from the model (bottom panel). Each frame corresponds to each category. The `Oth'
category does not follow any consistent pattern and is therefore not shown here. In each frame, a citation belt is formed by the lines $Q_1$ (green
line)
and $Q_3$ (blue line) which represent the first (10\% points lie below this line)
 and third  (10\% points lie above this line) quartiles
 of the data points respectively (i.e., effectively 80\% points are within citation belt), and the red line drawn within the
citation
belt represents the average behavior of all the profiles corresponding to that category. Top panel: for each category, one representative
citation profile (taken from real data) is shown at the middle of the belt (broken black line). Bottom panel: the color coding is similar to that of the top panel;
however, the broken lines are the results obtained from our model.
}\label{profile}
\end{figure*}

\section{Categorization of Citation Profiles}\label{sec:category}
Since the primary focus of our study is to analyze citation growth of a paper after publication, an in-depth understanding of how, after publication, the number
of citations of a paper varies over the years is necessary. We therefore conduct an exhaustive analysis of the citation
patterns of different papers present in our dataset. Some of the previous experimental results~\cite{asonam, garfield} show that the trend
of citations received by a paper after its publication date is not linear in general; rather there is a fast growth of citations within the
initial few years, followed by an exponential decay. This conclusion has been drawn mostly from the analysis of a small dataset of
publication archive. 
In this work, for an extensive analysis, we first take all the papers having at least 10 years of citation history, and consider maximum 20
years of their citation history.  This is followed by a series of data processing steps. First of all, to smoothen the time
series data points in the citation profile of a paper, we use five-years moving average filtering; then, we scale the data points by
normalizing them with the maximum value present in the time series (i.e, maximum number of citations received by the paper in some particular year);
finally, we run local peak detection algorithm\footnote{{\tiny The peak detection algorithm is available
in Matlab Spectral Analysis package - \url{http://www.mathworks.in/help/signal/ref/findpeaks.html}; we use `MINPEAKDISTANCE'=2 and
`MINPEAKHEIGHT'=0.75 and the default values for the other parameters.}} to detect peaks in the citation profile. In addition, we apply
the following two heuristics to specify peaks: (i) the height of a peak should be at least 75\% of the maximum peak-height, and (ii) two
consecutive peaks should be separated by more than 2 years; otherwise they are treated as a single peak. A systematic flowchart to detect each category is shown in Figure \ref{algo}.

\begin{figure}[h]
\centering
\includegraphics[width=\columnwidth]{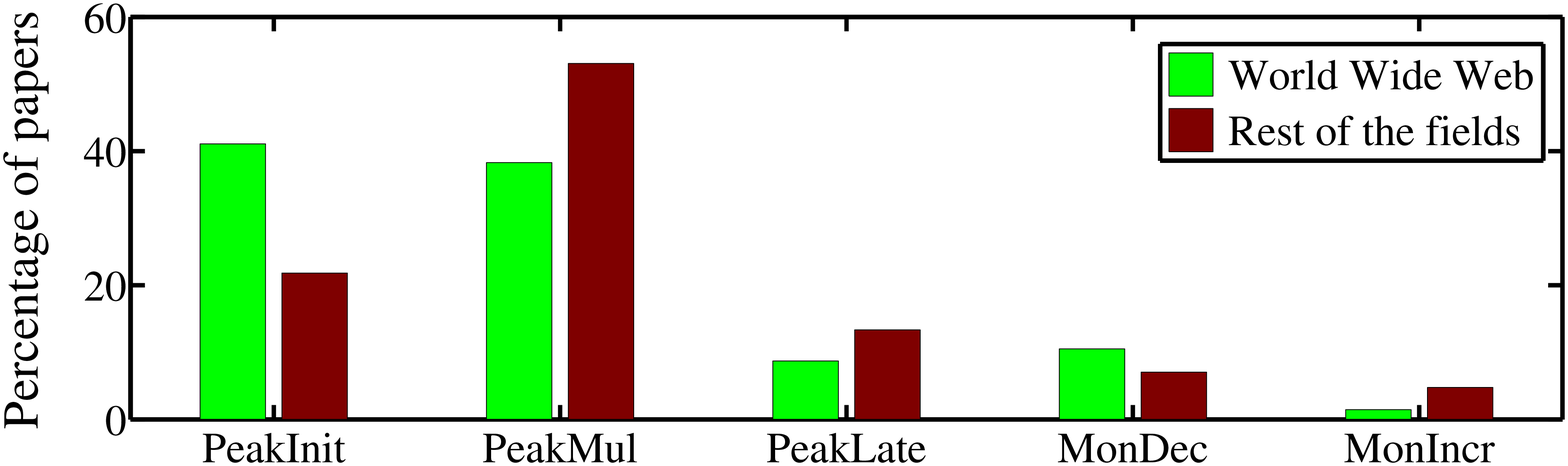}
\caption{(Color online) Percentage of papers in six categories for different research fields of computer science domain. For most of the fields, the pattern is similar, except World Wide Web.}\label{field}
\end{figure}

Remarkably, we notice that a major proportion of papers do not follow the traditional citation profile mentioned in the earlier studies
(see Figure \ref{belief});
rather there exist six different types of citation profiles of research papers based on the count and the position of peaks present in a profile. The definition of six types of citation profiles with the individual proportions in the entire dataset are give below:  \\
{\bf (i) PeakInit:} Papers whose citation count peaks within the first 5 years of  publication (but not in the first year)  followed by an
exponential decay (proportion: 25.2\%) (Figure~\ref{profile}(a)). \\
{\bf (ii) PeakMul:} Papers having multiple peaks at different time points of the citation profile (proportion: 23.5\%)
(Figure~\ref{profile}(b)).\\
{\bf (iii) PeakLate:} Papers having very few citations at the beginning and then a single peak after at least 5 years of
the
publication which is followed by an exponential decay in the citation count (proportion: 3.7\%) (Figure~\ref{profile}(c)).\\
{\bf (iv) MonDec:} Papers whose citation count peaks in the immediate next year of the publication followed by a monotonic decrease in the
number of citations (proportion: 1.6\%) (Figure~\ref{profile}(d)). \\
{\bf (v) MonIncr:} Papers having a monotonic increase in the number of citations from the very beginning of the year of publication till the
date of observation (i.e., it can be after 20 years of its publication) (proportion: 1.2\%) (Figure~\ref{profile}(e)). \\
{\bf (vi) Oth:} Apart from the above types, there exist a large number
of papers which on an average usually receive less than one citation per year. For these papers, the evidences are not
significant enough for
assigning them into one of the above categories, and, therefore, they remain as a separate group altogether (proportion: 44.8\%).

 The rich metadata information in the dataset further allows us to conduct a second level analysis of these categories for different
research fields of computer science domain. 
\begin{WrapText}
 $\bullet$ {\color{blue}WWW has the majority of papers in the PeakInit category.}\\
 $\bullet$ {\color{blue}Among all fields, Simulation and Computer Education have the highest proportion of papers in the MonDec category, while Bioinformatics and Machine Learning have the lowest.} \\
 $\bullet$ {\color{blue}Security and Privacy as well as Bioinformatics have the highest proportion of papers in the PeakLate category, while Simulation and WWW have the lowest.}
 \end{WrapText}
We measure the percentage of papers in different categories for each of the 24 research fields after filtering out all
the papers in the Oth category. Surprisingly, we notice that while for all other fields, maximum fraction of papers belong to the PeakMul category, 
for the field World Wide Web (WWW) this fraction is maximum in the PeakInit category (see Figure \ref{field}). The possible reason could be that since WWW is mostly a conference-based field of research, the papers in PeakInit category mostly dominate this field (see Section \ref{venue}).

\begin{figure}[h]
\centering
\includegraphics[width=\columnwidth]{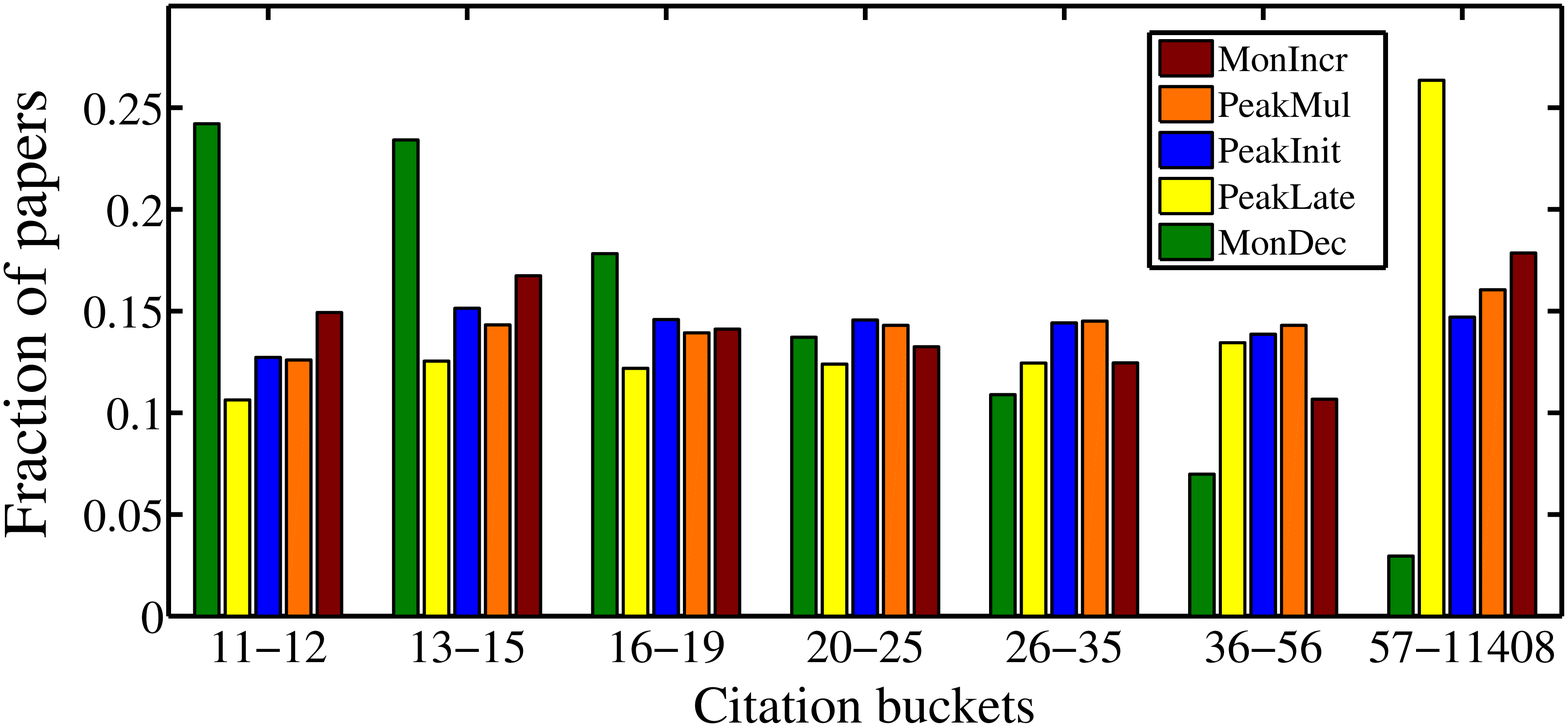}
\caption{(Color online) Contribution of papers from each category in different citation buckets. The entire range of citation value present in the dataset is
divided into seven buckets. In each bucket, the contribution of papers from a particular category is normalized by
the total number of papers
in that category. }\label{citation_bucket}
\end{figure}

\section{Contribution of Categories in Different Citation Ranges}
One of the fundamental aspects of analyzing scientific publications is to measure how acceptable they are to the research community. This
is 
often measured by the raw citation count -- the more citations an article receives from other publications, the more it is
assumed to be admired  by the
researchers and hence the more is the scientific impact \cite{bornmann}. In the current context, an interesting question is -- which among the six 
categories contains papers that are admired most in terms of citations. In order to answer this question, we conduct a systematic study --
the total citation range is divided
into four buckets (the citation ranges are: 11-12, 13-15, 16-19, 20-11408) such that each citation bucket would contain almost equal
number
of papers. For a deeper analysis of the highest citation range, we further divide the last bucket (20-11408) into four more ranges, thus
 obtaining seven buckets altogether.  
Then we measure the proportion of papers contributed by a particular category to a citation bucket (see Figure \ref{citation_bucket}).
Note that in each citation bucket, the number of
papers contributed by a category is normalized by the total number of papers belonging to that category. Therefore, this figure is
 a histogram of conditional probability distribution -- probability that a randomly selected paper falls in citation bucket $i$ given that
it belongs to
category $j$. The normalization is required in order to avoid population bias across different categories.
\begin{WrapText}
 $\bullet$ {\color{blue}Papers in PeakLate and MonIncr categories seem to receive maximum citations.}\\
 $\bullet$ {\color{blue}Papers in MonDec category mostly fall in the low citation region.} \\
 $\bullet$ {\color{blue}First few years' citation counts might not be a good indicator of the ultimate extent of acceptance of a paper in the scientific community.}
\end{WrapText}
We observe that 
the higher region of citation is mostly occupied by the papers in PeakLate and MonIncr categories followed by PeakMul and PeakInit. We
also notice that the MonDec category which has the minimum proportion in the last citation bucket shows a monotonic downward fall in the
fraction of papers as the citation range increases. These initial evidences present a general and non-intuitive interpretation of citation
profiles -- if a paper does not obtain high citations within the immediate few years after its publication, it does not necessarily mean that it will
continue to remain low impact all through its lifetime; rather in future its citation growth rate might
accelerate and it could indeed turn out to be a well accepted
paper in the scientific community. We further explain this behavior in the next section.

\begin{table}[h!]
\caption{Mean publication year Y (its standard deviation $\sigma(Y)$) and the percentage of papers in conferences and journals
for each category of citation profile.}\label{mean_year}
\scalebox{0.9}{
\begin{tabular}{|c|c|c|c|}\hline
Category & Mean publication & \% of conference & \% of journal \\
         &  year ($\sigma(Y)$) & papers & papers \\\hline
PeakInit &  1994 (5.19) & 64.35 & 35.65 \\\hline
PeakMul & 1991 (6.68) & 39.03 & 60.97 \\\hline
PeakLate &  1992 (6.54)& 39.89 & 60.11 \\\hline
MonDec & 1994 (5.44) & 60.73 & 39.27\\\hline
MonIncr & 1993 (7.36) & 25.26 & 74.74 \\\hline
\end{tabular}}

\end{table}

\vspace{-5mm}
\section{Characterizing Different Citation Profiles}

The rich metadata information of the publication dataset further allows us to understand the characteristic features of each of these six
categories at finer levels of detail.

\subsection{Influences of publication year and publication venues on the categorization} \label{venue}
One might raise an immediate question that this categorization might be influenced by the time (year) when the papers are published, i.e.,
the papers published earlier might be following the well-known behavior whereas the papers published recently might indicate a different
behavior. 
In
order to verify that the categorization is not biased by the publication time period, we measure the average year of publication of the
papers in each category. From the second column of Table \ref{mean_year}, we can conclude that the citation pattern of the
\begin{WrapText}
$\bullet$ {\color{blue}Due to the increasing popularity of conferences in an applied domain like computer science, the conference papers get
quick publicity within a few years after publication, which is also the reason for the rapid decay of their popularity. }\\
$\bullet$ {\color{blue}In contrast, journal papers usually take time to get published and also to get popularity, thus being mostly admired much later
 after publication. However, most of the journal  papers remain consistent in receiving citations even many years after their
publication.  }
\end{WrapText}
papers is not biased by the publication year since the average years roughly correspond to the same time period.
On the other hand, the mode of publication in conferences is significantly different from that of journals, and therefore the citation profiles of papers
published in these two venues are also expected to be different. 
To analyze the venue effect on the categorization, we measure
the percentage of papers published in journals vis-a-vis in conferences for each category as shown in the third and the fourth columns of
Table
\ref{mean_year} respectively. We observe that while most of the papers in PeakInit (64.35\%) and MonDec (60.73\%) categories are published in
conferences, papers belonging to PeakLate (60.11\%) and MonIncr (74.74\%) categories are mostly published in journals. 
Hence, if a publication starts receiving greater attention or citations at a later part of its lifetime,
it is more likely to be published in a journal and vice versa.

Another interesting point to be noted from these results is that although the existing formulation of the Journal Impact Factor
\cite{garfield} has been defined taking into consideration the citation profile as shown in Figure \ref{belief}, most of the journal
\begin{WrapText}
{\color{blue}
 The {\bf Impact Factor} \cite{garfield2} of a journal at any given time is the average number of citations received per paper published in that journal during the two preceding years.}
\end{WrapText}
papers which fall in PeakLate or MonIncr do not follow such a profile at all; at least for papers in PeakLate category, the metric does not
focus on the most-relevant time frame of the citation profile (mostly after first 5 years of  publication). 
In the light of the current
results, the appropriateness of the formulation of the bioliogaphic metrics such as Journal Impact Factor remain doubtful.

\begin{table}[!h]
\caption{(Color online) Confusion matrix representing the transition of categories due to the removal of self-citations. A value $x$ in the
cell $(i,j)$ represents that $x$ fraction of papers in category $i$ would have fallen in category $j$ if self-citations were absent in the
entire dataset. Note that, no row has been specified for Oth category because papers from this category
can never move to the other categories through deletion of citations.}\label{self_citation}
\scalebox{0.77}{
\begin{tabular}{|c|c|c|c|c|c|c|}\hline
 Category & PeakInit & PeakMul & PeakLate & MonDec & MonIncr & Oth \\\hline\hline
PeakInit & 0.72 & 0.10 & 0.03 & 0.01 & 0 & 0.15 \\\hline
PeakMul & 0.02 & 0.81 & 0.04 & 0 & 0.1 & 0.11 \\\hline
PeakLate & 0.01 & 0.06 & 0.86 & 0 & 0.01 & 0.06 \\\hline
MonDec & 0.05 & 0.14 & 0 & 0.41 &  0 & 0.35 \\\hline
MonIncr & 0 & 0.02 & 0.01 & 0.01 & 0.88 & 0.09 \\\hline
\end{tabular}}

\end{table}

 \begin{figure}[!h]
 \begin{center}
\includegraphics[width=\columnwidth]{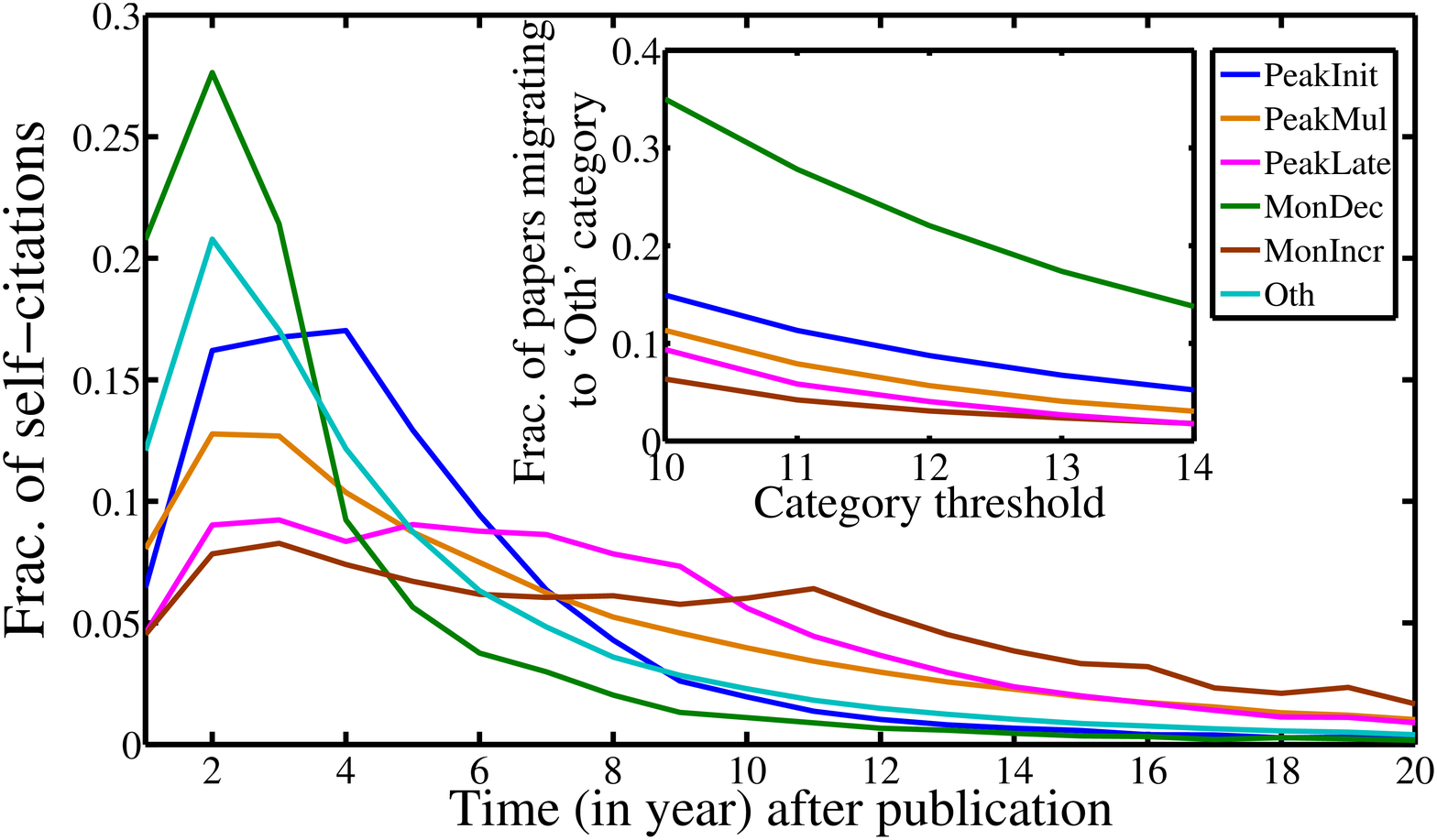}
\end{center}
\caption{(Color online) Faction of self-citations per paper in different categories over different time periods after
publication; (inset) fraction of papers in each category migrating to Oth category due to removal of self-citations assuming different category thresholds.}\label{self-cite}
\end{figure}

\begin{figure*}[!ht]
\centering
\scalebox{0.3}{
\includegraphics{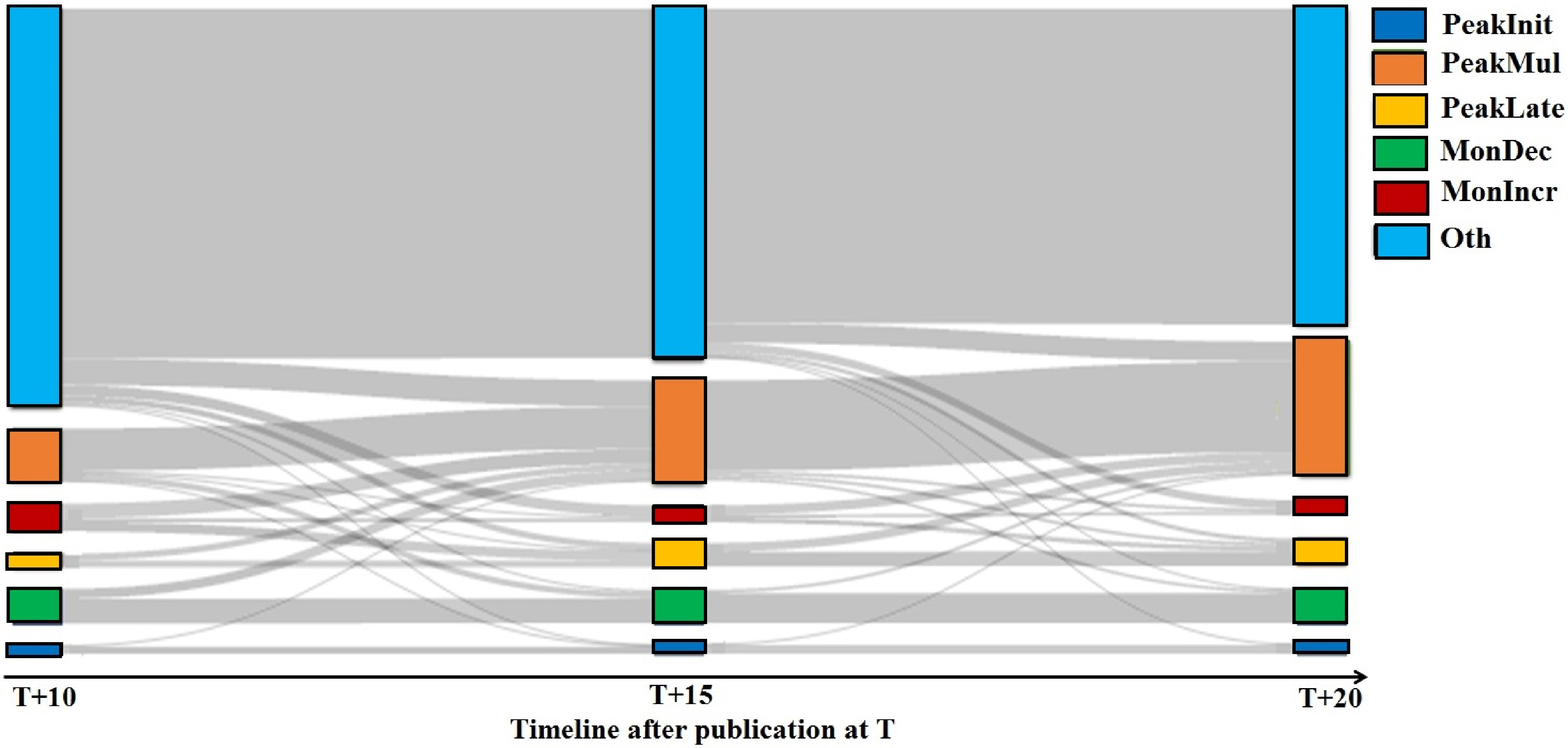}}
\caption{(Color online) Alluvial diagram representing  the evolution of papers in different categories and the flows between the categories
at time $T+10$, $T+15$ and $T+20$. The colored blocks correspond to different categories. The size of the block indicates the number of
papers in that category, and the shaded waves joining the regions represent flow of papers between the regions, such that the width of the
flow corresponds to the fraction of papers. The total width of incoming flows is equal to the width of the corresponding
region.}\label{alluvial}
\end{figure*}

\subsection{Effect of self-citation on the categorization}
Another factor that often affects citation rate is self-citation \cite{fowler07citation}.
 Self-citation can inflate the perception of an article's or a
scientist's scientific impact, particularly when an article has many authors, increasing the possible number of self-citations
\cite{Sala, Schreiber}, and thus there have been calls to remove self-citations from citation rates \cite{Schreiber}. 
We also conduct a similar experiment to notice the effect of self-citation on the categorization of citation profiles.
\begin{WrapText}
 $\bullet$ {\color{blue}Authors tend to cite their own papers within 2-3 years of their publications in order to increase the visibility of their work to the audience.}\\
 $\bullet$ {\color{blue}MonDec and PeakInit categories (i.e., mostly the conference papers) are highly affected by the self-citation.}\\
 $\bullet$ {\color{blue}The self-citation is usually used in initial periods of the publication by the authors in an attempt to increase the visibility of their publications in the scientific community.}
\end{WrapText}
Essentially, we first 
dispose the citation from the dataset if the citing and the cited papers have at least one author in common, and then measure what fraction
of papers in each category migrate to some other category due to this disposal. Table \ref{self_citation} presents a confusion matrix where
labels in the rows and the columns represent the categories before and after removing self-citations respectively. We observe that papers in MonDec are vastly affected
by the self-citation phenomenon. Around 35\% of papers in MonDec would have been in the `Oth' category had it not been due to
the self-citations. However, one might argue that this might be the artifact of the thresholding that we impose (see Section \ref{sec:category}) to categorize papers (papers receiving less than or equal to 10 citations in first 10 years after publication are considered to be in `Oth' category). In order to verify the effect of thresholding on the inter-category migration after removing self-citations, we vary the category threshold from 10-14 and plot in Figure \ref{self-cite} (inset) the fraction of papers in each category migrating to Oth category due to the disposal of self-citations. The result agrees with the observation noted in Table \ref{self_citation}; MonDec category is mostly affected by self-citations, followed by PeakInit, PeakMul and PeakLate. This indicates that the effect of self-citations is due to the inherent characteristics of each category, rather than due to the predefined threshold setting of the category boundary.

In Figure \ref{self-cite}, we show  how the self-citations are distributed across different time periods for individual categories. For each category, we first aggregate all the self-citations
and plot the fraction of self-citations over the time after publication. As expected, for MonDec category we observe that most number of
self-citations are ``farmed'' within 2-3 years after publication. Similar observations hold for both the PeakInit and Oth categories. Note that,
as observed earlier, PeakInit and MonDec categories are mostly found to be conference papers. Therefore, we can conclude that conference
papers tend to be
mostly affected by self-citations. However, the characteristics of the highly-cited categories such as MonIncr and PeakLate are mostly
consistent throughout the years which show that these categories are less dependent on self-citations.

\section{Analyzing Stability of Different Categories}
The number of citations for a paper changes over time depending on its long/short lasting effect on the scientific community which in turn
might
change the shape of the citation profile. Therefore, studying the temporal evolution of each citation profile can help us 
understand the stability of the categories individually. Since, we know the category of those papers that have at least 20 years of citation
history, for each such paper we further analyze how the shape of the profile evolves through these 20 years timeline. Essentially, after
publication
of a paper at time $T$, we identify its category at time $T+10$, $T+15$ and $T+20$ based on the heuristics discussed earlier. We hypothesize that a stable citation category tends to maintain its shape throughout the entire timeline. 
The
colored blocks of the alluvial diagram \cite{Rosvall} in Figure \ref{alluvial}  correspond to the different
\begin{WrapText}
 $\bullet$ {\color{blue}Papers in Oth category often shift to MonIncr category by acquiring more citations in the later time period.}\\
 $\bullet$ {\color{blue}If a paper falls in either MonDec or PeakInit category earlier, the likelihood that it would shift to some other category is less.} 
\end{WrapText}
categories for three different timestamps.  We observe that apart from the Oth category which has a
major
proportion of papers, MonDec seems to be the most stable, which is followed by PeakInit. However, papers which are assumed to fall in Oth category quite often turn out to be MonIncr
papers in the later time periods. This analysis indeed demonstrates a systematic approach to unfold the transition from one category to
another  taking place in scientific research with the increase of citations.
%PG: Last line on its own doesn't look good. DONE

\begin{figure*}[t!]
\centering
\scalebox{0.27}{
\includegraphics{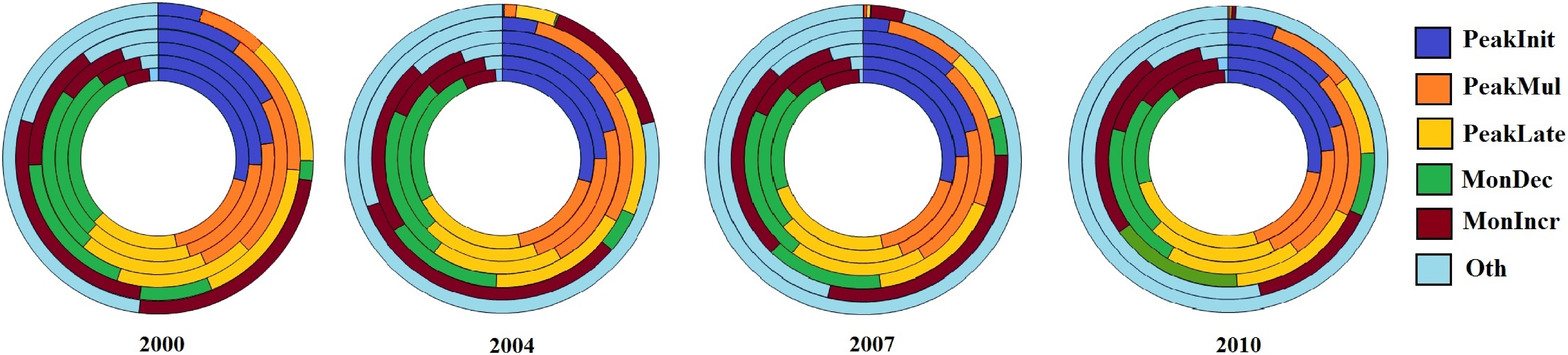}}
\caption{(Color online) Multi-level pie chart for year 2000, 2004, 2007 and 2010 showing the composition of each of the categories in different $k^s$-shell
regions; where the colors represent different categories and the area covered by each colored region in each $k^s$-shell denotes the
proportion of papers in the corresponding category occupied in that shell. The innermost shell is the core region and the outermost shell
is the periphery region. For better
visualization, we divide the total number of shells identified from the citation network in each year into six broad
shells; thus the core-periphery structure in each year has six concentric layers.}\label{k-shell}
\end{figure*}

\section{Core-periphery Analysis}
\if{0}
Although Figure \ref{citation_bucket} provides the impact of different categories in terms of raw citation count,  it neither indicates the
significance of the papers in each category forming the core of the network nor gives us any information regarding the temporal evolution of
the structure. For a better and more detailed understanding, we perform $k$-core analysis \cite{harris2008, seidman1983network,
ShaiCarmi07032007, Kumar} of the evolving citation network by decomposing the network for each year into its $k^s$-shells (see Materials
and Methods), such that an inner shell index of a paper reflects a central position in the core of the network. For better
visualization in Figure \ref{k-shell}, we divide the total number of shells identified from the citation network in each year into six broad
shells; thus the core-periphery structure in each year has six concentric layers. The idea is to show how the papers in each category
(identified at the year 2000) migrates from one shell to another after getting citations in the next 10 years. It also allows us to
observe how persistent a category is in a particular shell.

In Figure \ref{k-shell}, we notice that the majority of papers in the Oth category lie in the periphery and its proportion in the periphery
increases
over time, which indicates that the papers in this category are becoming increasingly less popular in time. PeakMul category exhibits a
decent presence in all the shells at 2000; over time gradually it leaves the peripheral region and mostly occupies the two innermost
shells. 
PeakInit and MonDec shows almost similar behavior with a major proportion of papers in inner cores in the initial year but
gradually shifting towards peripheral regions. Their placement in the inner core might be due to the fact that they are associated to high
ranked authors and include self citations. Hence due to the highly
connected and cited neighborhoods, some of them make a position in the central core. On the other
hand, MonIncr and PeakLate show
expected behavior with their proportion increasing towards inner shells over time indicating their rising relevance as time progresses.

This study helps us identify temporal evolution of the importance of different categories in terms of how each of
them contributes to the central position of the citation network. 

\fi

Although Figure \ref{citation_bucket} indicates the impact of different categories in terms of raw citation count,  it neither unfolds the
significance of the papers in each category forming the core of the network nor gives us any information regarding the temporal evolution of
the structure. For a better and more detailed understanding, we perform $k$-core analysis \cite{ShaiCarmi07032007, harris2008} of the evolving citation network by decomposing the network for each year into its $k^s$-shells such that an inner shell index of a paper reflects a central position in the core of the network. 

 We construct different aggregated citation networks in different years -- 2000, 2004, 2007 and 2010
such that citation network constructed in the year $Y$ contains the induced subgraph of all papers published at or before $Y$. Then for each
such network, we run the following methods: we start by recursively removing nodes that have a single inward link until no such nodes remain in the network. These nodes form the $1$-shell of the network ($k^s$-shell index $k^s = 1$). Similarly, by recursively removing all nodes with degree 2, we get the $2$-shell. We continue increasing $k$ until all nodes in the network have been assigned to one of the shells. The union of all the shells with index greater than or equal to $k^s$ is called the $k^s$-core of the network, and the union of all shells with index smaller or equal to $k^s$ is
the $k^s$-crust of the network. The idea is to show how the papers in each category (identified in the year 2000) migrates from one shell to another after getting citations in the next 10 years. It also allows us to observe how persistent a category is in a particular shell.

In Figure \ref{k-shell}, we notice that the majority of papers in the Oth category lie in the periphery and its proportion in the periphery
increases over time which indicates that the papers in this category are becoming increasingly less popular over time. PeakMul category gradually leaves
the peripheral region over time and mostly occupies the two innermost shells. PeakInit and MonDec show almost similar behavior with a major
proportion of papers in inner cores in the initial year but gradually shifting towards peripheral regions. On the other
hand, MonIncr and PeakLate show expected behavior with their proportion increasing in the inner shells over time indicating their rising
relevance as time progresses. This study helps us identify temporal evolution of the importance of different categories in terms of how each
of
them contributes to the central position of the citation network.

\section{Modeling Citation Profiles Using Dynamic Growth Model}
There has been extensive research done in developing growth models to explain the evolution of citation networks \cite{journals, Sen2005}.
Models like Barab\'asi-Albert \cite{Albert2002, Barabasi99}, Price \cite{Price} etc.  attempt to generate scale-free networks using
preferential attachment mechanism. Most of these works seek to explain the emergence of mainly the degree distribution of the network.
In
this paper, we propose a novel dynamic growth model to synthesize the citation network with the aim of reproducing the citation
categories obtained from the
real-world
dataset. To the best of our knowledge, this model is the first of its kind which takes into account two major components,
namely preferential
attachment \cite{Albert2002} and aging \cite{Gingras,RePEc:eee} in order to mimic the real-world
citation profiles.

% In order to avoid the cold-start problem \cite{Schein:2002}, we include in the model certain information of the initial few years (nodes
% and their categories, year of publication, citation
% edges) from the real dataset to perform bootstrapping.

We use the following
distributions as inputs to the model for a fair comparison with the real-world citation profiles: distribution of the number of papers over
the
years (to determine the
influx of papers into the system at each time step), reference distribution (to determine the outward citations that would emanate from an
incoming node). 
\begin{WrapText}
 $\bullet$ {\color{blue}{\bf Preferential attachment} means that the more connected a node is, the more likely it is to receive new connections. Preferential attachment is an example of a positive feedback cycle where initially random variations (one node initially having more links or having started accumulating links earlier than another) are automatically reinforced, thus greatly magnifying differences.}\\
 $\bullet$ {\color{blue}In many growing networks, the {\bf age} of the nodes plays an important role in deciding the attachment probability of the incoming nodes. For example, in a citation network, very old papers are seldom cited while recent papers are usually cited at a higher frequency. }
\end{WrapText}
At each time step (corresponding to a particular year), a number of nodes (papers) are selected with the outdegree
(references) for each of them determined preferentially from the reference distribution. 
Then the vertex
is assigned preferentially to a certain category based on the size of the categories (number of papers in the categories) at that time
step. For determining the other end point of each edge associated with the incoming node,  we first select a category preferentially based
on
the in-citation information of each category, and then within the category we select a node preferentially based on its {\em attractiveness}.
The {\em attractiveness} of a node (paper) is determined by the time elapsed since its publication (aging) and the number of citation
accumulated till that time (preferential attachment).  Note that, the formulation of the attractiveness in our model also varies for
different categories (see SI text).

We observe a remarkable resemblance between the real-world citation profiles and those obtained from the model as shown in Figure
\ref{profile} (bottom panel: $(a1)$-$(e1)$).  Each frame of this figure contains three lines depicting first quartile (10\% points lie below
this
line), third quartile (10\% points lie above this line) and the mean behavior. We also compare the in-degree distributions obtained from the model and from the real dataset for different categories and observe a significant resemblance
(see SI text). Hence our model presents
a holistic view of the evolution of a citation network over time and the intra- and inter-category interactions that 
account for the observable properties of the real-world system.

%% file: discussion.tex
\section{Discussion}
The collection of massive computer science bibliographic dataset allows us to conduct such exhaustive analysis of citation profiles of
individual papers and to derive six predominant categories that remained unobserved in the literature so far. At the microlevel, this paper
provides, for example, a set of new approaches to characterize each individual category as well as to study the dynamics of their
evolution over time. Finally, leveraging on these behavioral signatures we are able to design a novel dynamic model to
synthesize the real-world network evolving over time. This model in turn intrinsically unfolds the citation patterns of different
categories, which show a significant resemblance to that obtained from the real data.

This paper thus offers a necessary first step towards reformulating the existing quantifiers available in Scientometrics that
should leverage the signature of different citation patterns in order to formulate robust measures. Moreover, we believe that a systematic
machine learning model of the behavior of different citation patterns has the potential to enhance the standard research methodology in
this area which includes topics like discovering missing links in citation networks \cite{Clauset}, early prediction of citations of
scientific articles
\cite{Yan:2012}, predicting high-impact and seminal papers \cite{McNamara}, recommending scientific articles \cite{Bogers:2008} etc.

In future, we plan to extend our study on the datasets of other domains such as
physics and biology to verify the universality of such categorizations. Moreover, we are keen to understand the micro-level
dynamics controlling the behavior of PeakMul category which is significantly different from the others. One initial observation towards this
direction is that PeakMul behaves like the intermediary between PeakInit and PeakLate categories (see SI text). In future,
we would like to conduct a detailed  analysis to understand different characteristic features particularly for the PeakMul category.